\documentclass[final]{svjour2}
\newcommand{\subparagraph}{}
\usepackage{newtxtext,newtxmath}
\usepackage{graphicx}
\usepackage{rotating}
\usepackage{amssymb}
\usepackage{mathptmx}
\usepackage[numbers]{natbib}
\usepackage{lineno}
\usepackage{hyperref}
\hypersetup{colorlinks,allcolors=black}
\usepackage{xcolor}
\usepackage{tabularx}
\usepackage{tabu}
\usepackage{soul}
\usepackage{textcomp}
\usepackage{subcaption}

\newcommand{\bigO}{$\mathcal{O}$} 
\newcommand{\parthz}{$\frac{\mathrm{pA}}{\sqrt{\mathrm{Hz}}}$}
\makeatletter
\journalname{Journal of Low Temperature Physics}
\begin{document}

\title{Simons Observatory Microwave SQUID Multiplexing Readout - Cryogenic RF Amplifier and Coaxial Chain Design}

\author{Mayuri Sathyanarayana Rao$^{\dagger}$\textsuperscript{1}\kern-1.5pt \and
Maximiliano Silva-Feaver$^{\ddagger}$\textsuperscript{2}\kern-1.5pt \and
Aamir Ali\textsuperscript{3}\kern-1.5pt \and
Kam Arnold\textsuperscript{2}\kern-1.5pt \and
Peter Ashton\textsuperscript{1,3,4}\kern-1.5pt \and
Bradley J. Dober\textsuperscript{5}\kern-1.5pt \and 
Cody J. Duell\textsuperscript{6}\kern-1.5pt \and
Shannon M. Duff\textsuperscript{5}\kern-1.5pt \and 
Nicholas Galitzki\textsuperscript{2}\kern-1.5pt \and 
Erin Healy\textsuperscript{7}\kern-1.5pt \and 
Shawn Henderson\textsuperscript{8}\kern-1.5pt \and 
Shuay-Pwu Patty Ho\textsuperscript{7}\kern-1.5pt \and
Jonathan Hoh\textsuperscript{9}\kern-1.5pt \and
Anna M. Kofman\textsuperscript{10}\kern-1.5pt \and
Akito Kusaka\textsuperscript{1,3,4}\kern-1.5pt \and
Adrian T. Lee\textsuperscript{1,3}\kern-1.5pt \and 
Aashrita Mangu\textsuperscript{3}\kern-1.5pt \and
Justin Mathewson\textsuperscript{9}\kern-1.5pt \and
Philip Mauskopf\textsuperscript{9}\kern-1.5pt \and
Heather McCarrick\textsuperscript{7}\kern-1.5pt \and
Jenna Moore\textsuperscript{9}\kern-1.5pt \and
Michael D. Niemack\textsuperscript{6}\kern-1.5pt \and 
Christopher Raum\textsuperscript{3}\kern-1.5pt \and 
Maria Salatino\textsuperscript{11}\kern-1.5pt \and
Trevor Sasse\textsuperscript{3}\kern-1.5pt \and 
Joseph Seibert\textsuperscript{2}\kern-1.5pt \and
Sara M. Simon\textsuperscript{12}\kern-1.5pt \and
Suzanne Staggs\textsuperscript{7}\kern-1.5pt \and
Jason R. Stevens\textsuperscript{6}\kern-1.5pt \and 
Grant Teply\textsuperscript{2}\kern-1.5pt \and
Robert Thornton\textsuperscript{13}\kern-1.5pt \and
Joel Ullom\textsuperscript{5}\kern-1.5pt \and 
Eve M. Vavagiakis\textsuperscript{6}\kern-1.5pt \and
Benjamin Westbrook\textsuperscript{3}\kern-1.5pt \and 
Zhilei Xu\textsuperscript{10}\kern-1.5pt \and
Ningfeng Zhu\textsuperscript{10}\kern-1.5pt
}

\institute{\footnotesize
\noindent\textsuperscript{1}Lawrence Berkeley National Laboratory, Berkeley, CA, USA \\
\noindent\textsuperscript{2}University of California San Diego, San Diego, CA, USA \\
\noindent\textsuperscript{3}University of California Berkeley, Berkeley, CA, USA \\
\noindent\textsuperscript{4}Kavli IPMU, Kashiwa, Chiba, Japan \\
\noindent\textsuperscript{5}NIST, Boulder, CO, USA \\
\noindent\textsuperscript{6}Cornell University, Ithaca, NY, USA \\
\noindent\textsuperscript{7}Princeton University, Princeton, NJ, USA \\
\noindent\textsuperscript{8}SLAC National Accelerator Laboratory, Menlo Park, CA, USA \\
\noindent\textsuperscript{9}Arizona State University, Tempe, AZ, USA \\
\noindent\textsuperscript{10}University of Pennsylvania, Philadelphia, PA, USA\\
\noindent\textsuperscript{11}KIPAC/Stanford University, Standford, CA, USA \\
\noindent\textsuperscript{12}University of Michigan, Ann Arbor, MI, USA \\
\noindent\textsuperscript{13}West Chester University of Pennsylvania, West Chester, PA, USA \\
\email{$^{\dagger}$msrao@lbl.gov, $^{\ddagger}$msilvafe@ucsd.edu}
}

%\author{Max Silva-Feaver$^{a}$ \and 
%Mayuri Sathyanarayana Rao$^{b,c}$}

%\institute{
%$^{a}$ Department of Physics, University of California San Diego, LaJolla, CA, USA
%\\$^{b}$ Physics Division, Lawrence Berkeley National Laboratory, Berkeley, CA, USA
%\\$^{c}$ Department of Physics, University of California, Berkeley, CA, USA
%\\ \email{$^{\dagger}$msrao@lbl.gov, $^{\ddagger}$msilvafe@ucsd.edu}
%}
\maketitle 
\begin{abstract}
The Simons Observatory (SO) is an upcoming polarization-sensitive Cosmic Microwave Background (CMB) experiment on the Cerro Toco Plateau (Chile) with large overlap with other optical and infrared surveys (e.g., DESI, LSST, HSC). To enable the readout of \bigO(10,000) detectors in each of the four telescopes of SO, we will employ the microwave SQUID multiplexing technology. With a targeted multiplexing factor of \bigO{(1,000)}, microwave SQUID multiplexing has never been deployed on the scale needed for SO.  Here we present the design of the cryogenic coaxial cable and RF component chain that connects room temperature readout electronics to superconducting resonators that are coupled to Transition Edge Sensor bolometers operating at sub-Kelvin temperatures. We describe design considerations including cryogenic RF component selection, system linearity, noise, and thermal power dissipation. %, SO will serve as an important test case for future large detector count experiments such as CMB-S4 \citep{CMB-S4}. 
\end{abstract}\\
\textbf{Keywords:} SQUID -- multiplexing -- readout -- CMB -- instrument design 
\section{Introduction}
\begin{figure}[]
    \centering
    \includegraphics[scale=0.3]{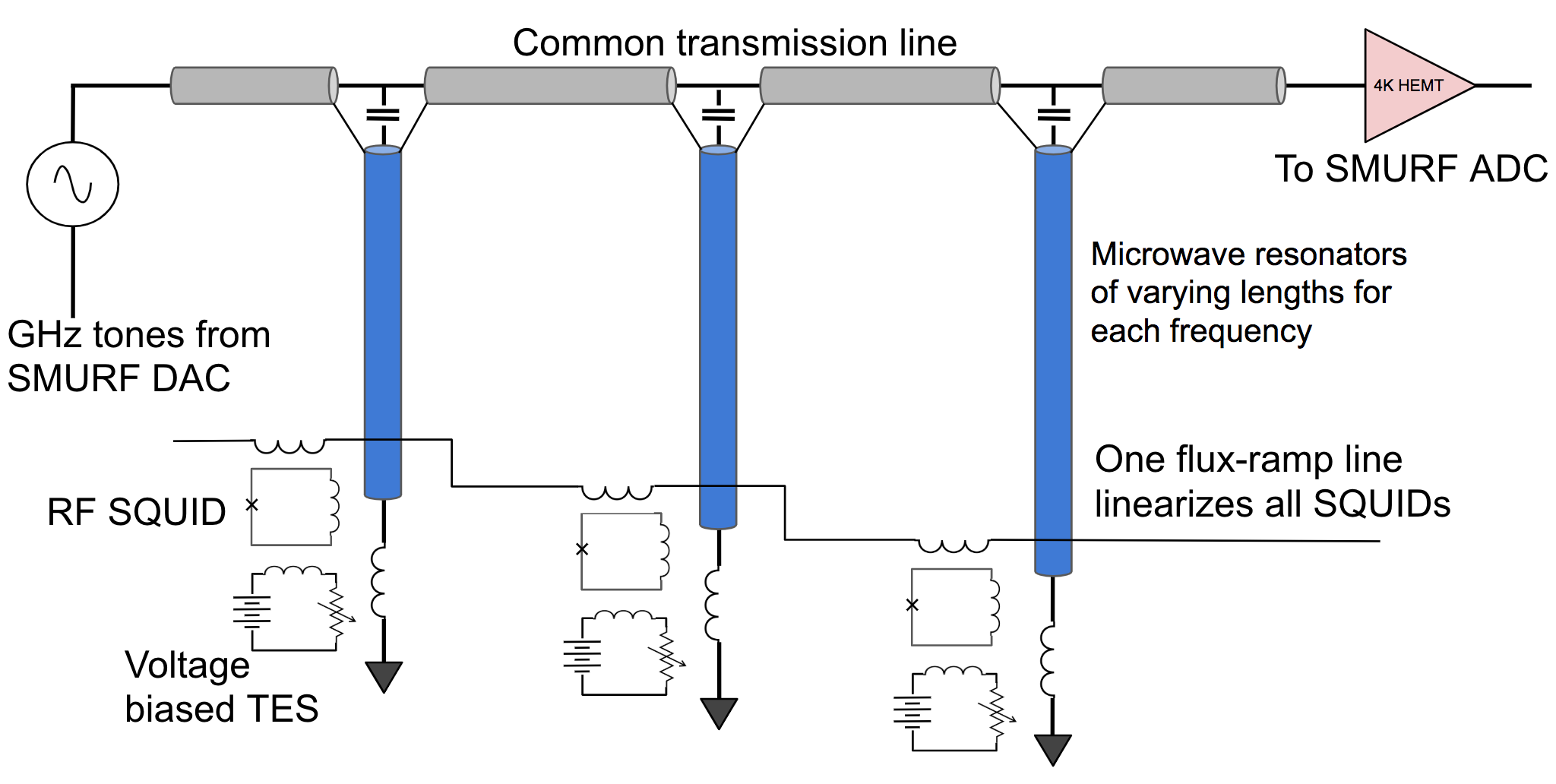}
    \caption{Schematic of the $\mu$-mux TES readout. Resonators, each designed to have a different frequency, are capacitively coupled to a single microwave transmission line. Each resonator is in-turn coupled inductively to a DC voltage-biased TES detector and a dissipationless RF SQUID. The SQUIDs are all linearized using a common flux-ramp line. Thus several detectors are read out using a pair of coaxial cables. The targeted multiplexing factor for SO is \bigO{(1000)}}.
    \label{fig:mumuxschem}
\end{figure}

The Simons Observatory (SO) \citep{SO_instrument} is an upcoming polarization-sensitive Cosmic Microwave Background (CMB) experiment in the Atacama desert in Chile. SO is building three small aperture telescopes (SATs) and one large aperture telescope (LAT). The SATs \cite{SAT_2019} target degree-scale B-mode (divergence-free) polarization of the CMB. The LAT targets small angular scale science (e.g., a possible discovery of new light relic particles, measurement of neutrino mass, measurement of the number of relativistic species, tests for deviation of cosmological constant, studies of gravitational lensing of the cosmic microwave background,  to name a few) \cite{SO_science}. To achieve the sensitivity required to meet SO science goals, we need to increase the number of detectors per receiver significantly. This calls for highly multiplexed readout to work within tight space and thermal loading constraints and over an order of magnitude increase in the multiplexing factor from existing experiments ($\sim$70x) \cite{SPT3G_2019_fmux,Adv_ACTPol_2016}. We have chosen to implement Microwave SQUID Multiplexing ($\mu$-mux) as the readout technology \citep{mumux_cmb,Matesthesis}, with a targeted multiplexing factor of \bigO({1,000}). $\mu$-mux combines the high multiplexing factor of Microwave Kinetic Inductance Detectors (MKIDs) \cite{McHugh2012} with the sensitivity demonstrated by millimeter-wave antenna-coupled DC-biased Transition Edge Sensor (TES) bolometers. The $\mu$-mux readout scheme is shown schematically in Fig. \ref{fig:mumuxschem}. It consists of a comb of resonators, spread over 4-8~GHz, coupled to a common transmission line, where each resonator is coupled to a dissipationless radio-frequency superconducting quantum interference device (RF-SQUID). The RF-SQUID is in turn inductively coupled to a voltage-biased TES. All SQUIDs are linearized using a low-frequency flux ramp modulation line \cite{Matesthesis}. The cold readout components for SO (resonators and SQUIDs) are developed by NIST (National Institute of Standards and Technology). These components along with detectors are packaged into Universal $\mu$-mux Modules (UMMs) \cite{HM_SO_2019,Li_SO_2019}. SLAC Superconducting Microresonator RF (SMURF) electronics \cite{smurf},  from SLAC National Accelerator Laboratory, serve as the room temperature readout electronics. 
\\\\
No other CMB experiment to date has targeted the high multiplexing factors that Simons Observatory (SO) aims to achieve. Progress with warm electronics (FPGAs) and cold readout design ( superconducting microwave resonators, SQUIDs) are essential to achieve this target. However, being able to effectively couple cold readout electronics at the detectors to warm electronics is non-trivial. Challenges include mechanical complexity, thermal loading, system linearity, and noise, to name a few. This paper presents a forward looking approach to RF wiring design for SO. The main design challenges are from space and linearity constraints. The space constraints are from the design of cryogenic chambers that are optimized for optical throughput with little space available for the readout components and wiring. Driving the microwave resonators at high powers to minimize two-level system (TLS) noise while simultaneously reading out $\mathcal{O}$(1,000) resonators pushes the limits of commercially available cryogenic low-noise amplifiers, makes linearity a critical design challenge. 

\section{SO RF readout wiring overview}
\begin{figure}[t]
    \centering
    \begin{subfigure}[b]{\textwidth}
    \includegraphics[width=\textwidth,height=2.5cm]{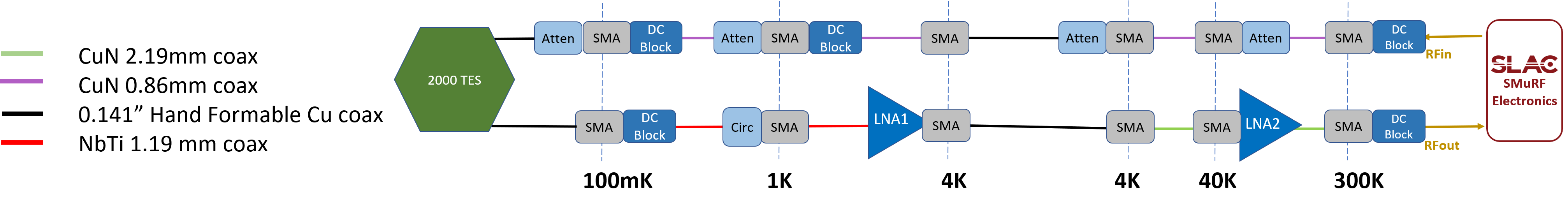}
    \caption{SO wiring design}
    \label{fig:SO_wiring}
    \end{subfigure}
    \begin{subfigure}[b]{0.5\textwidth}
        \centering
        \includegraphics[width=\textwidth]{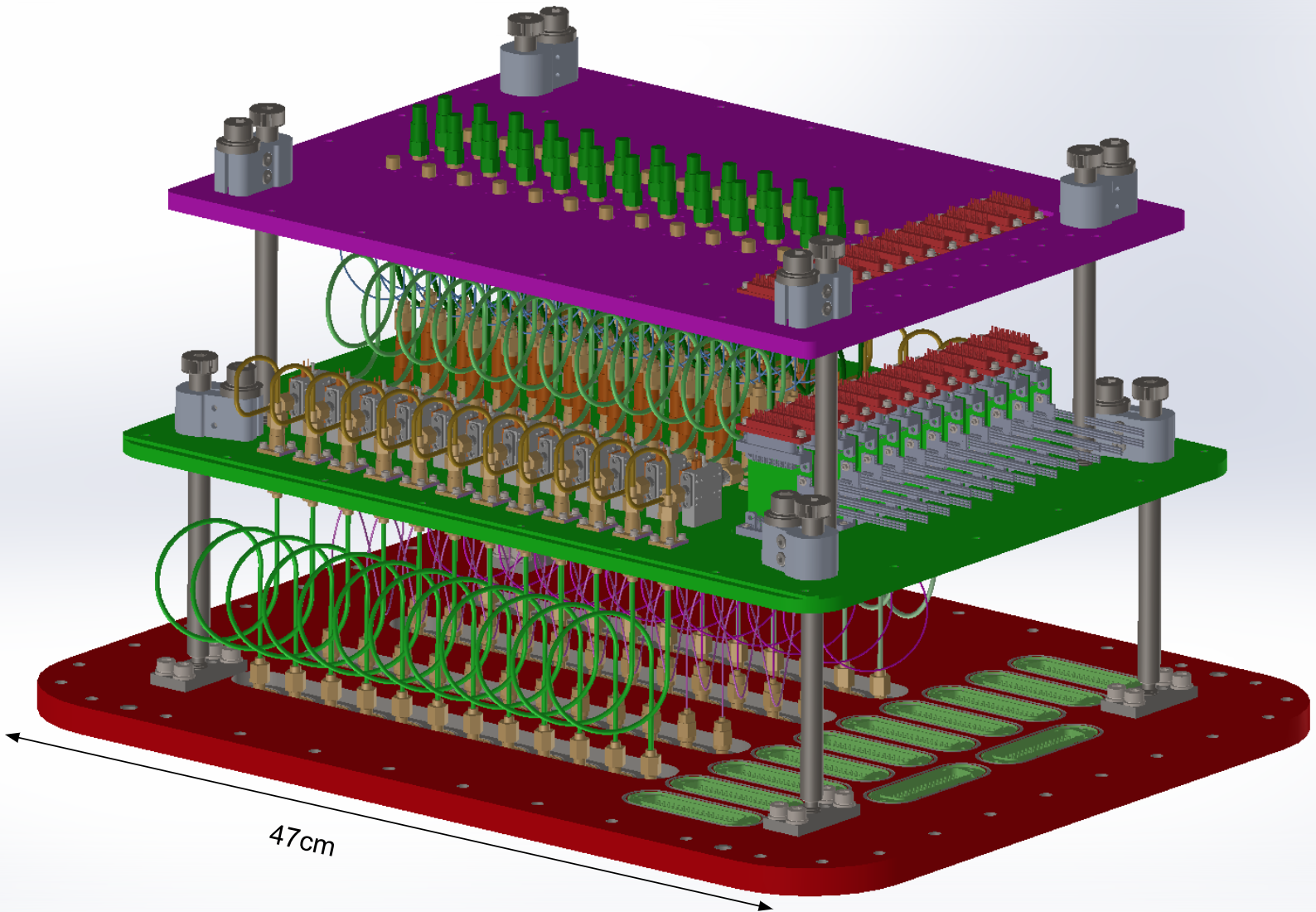}
        \caption{URH design}
        \label{fig:URH_design}
    \end{subfigure}
    \hfill
    \begin{subfigure}[b]{0.4\textwidth}
        \centering
        \includegraphics[width=\textwidth]{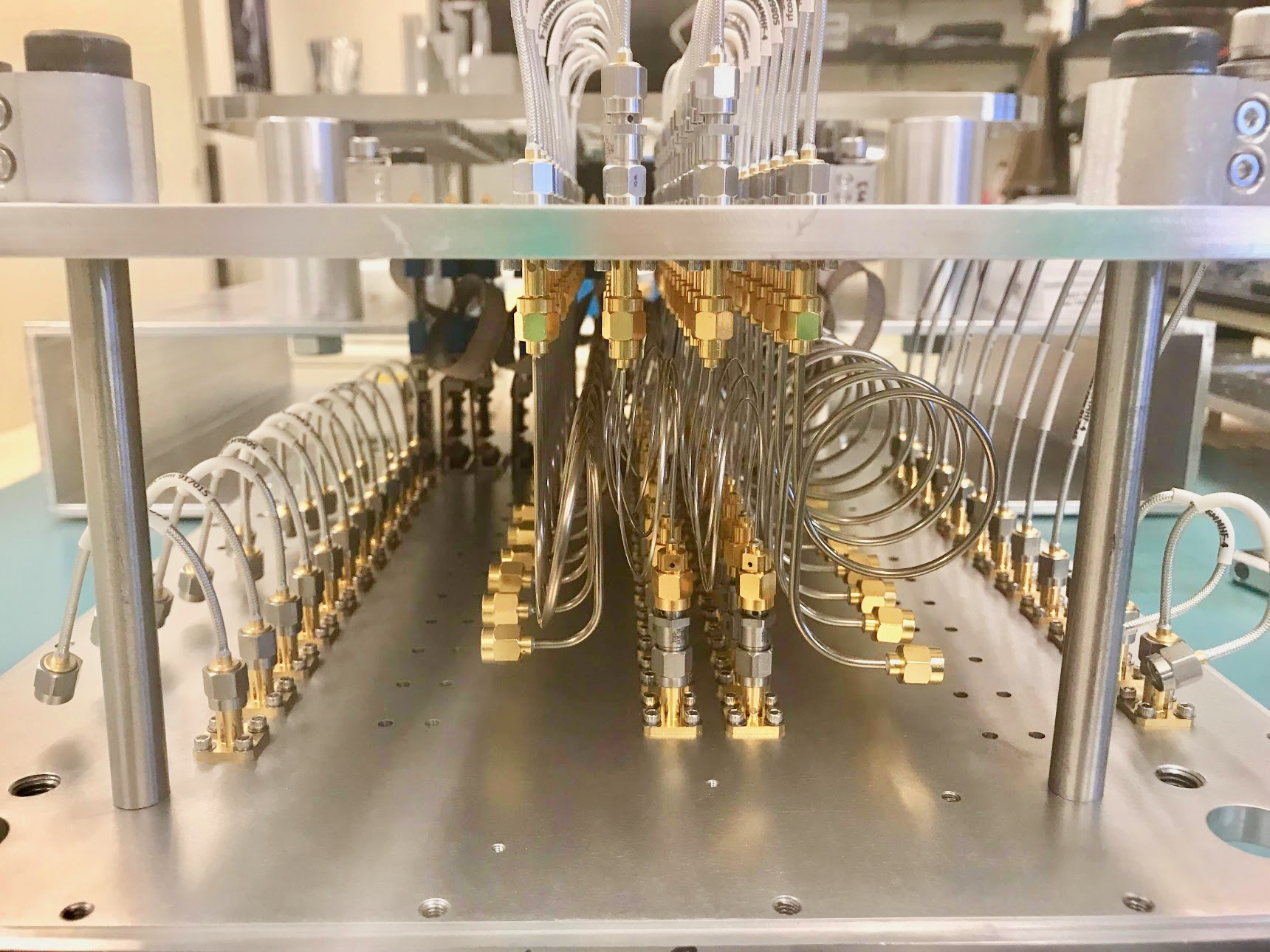}
        \caption{Wiring in URH during assembly}
        \label{fig:URH_picture}
    \end{subfigure}
    \quad
    \begin{subfigure}[b]{0.4\textwidth}
        \centering
        \includegraphics[width=\textwidth]{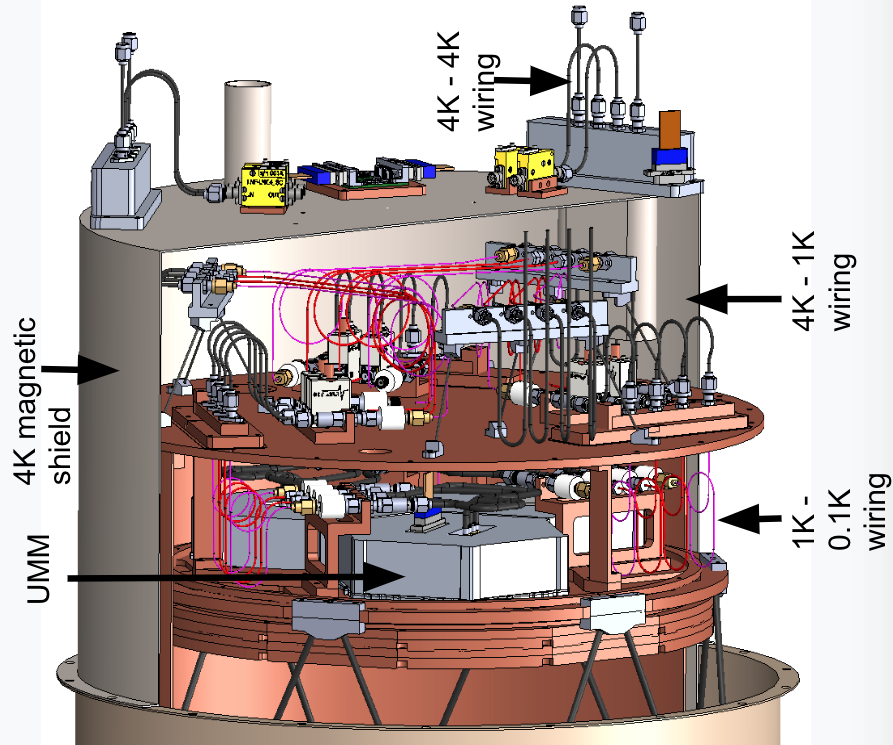}
        \caption{Cross section view of LATR cold wiring design}
        \label{fig:LATR_design}
    \end{subfigure}
    \hfill
    \begin{subfigure}[b]{0.4\textwidth}
        \centering
        %\begin{turn}{180}
        \includegraphics[width=\textwidth]{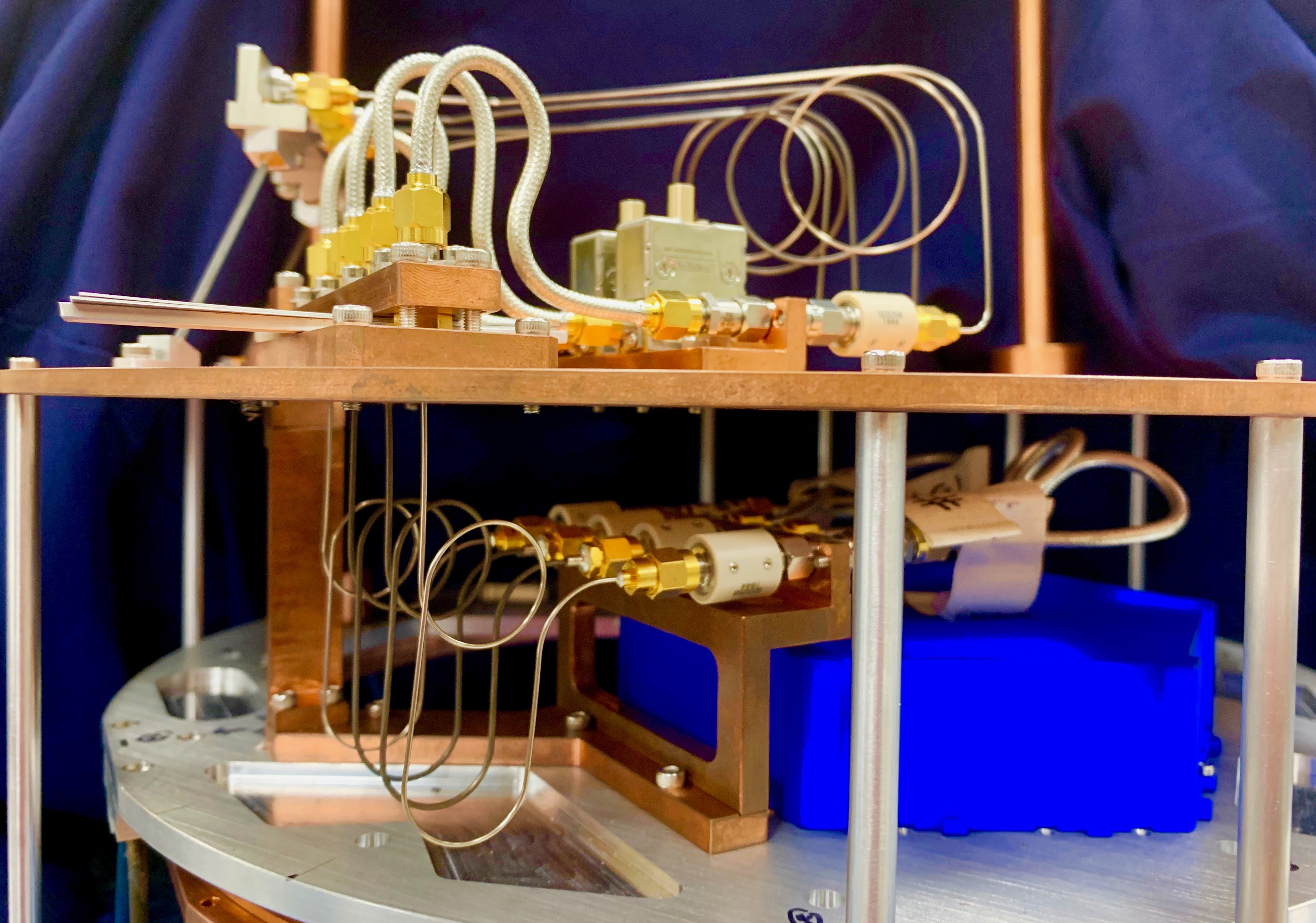}
        %\end{turn}
        \caption{LATR cold readout wiring prototype}
        \label{fig:LATR_picture}
    \end{subfigure}
    \quad
    \begin{subfigure}[b]{0.5\textwidth}
        \centering
        \includegraphics[width=0.9\textwidth]{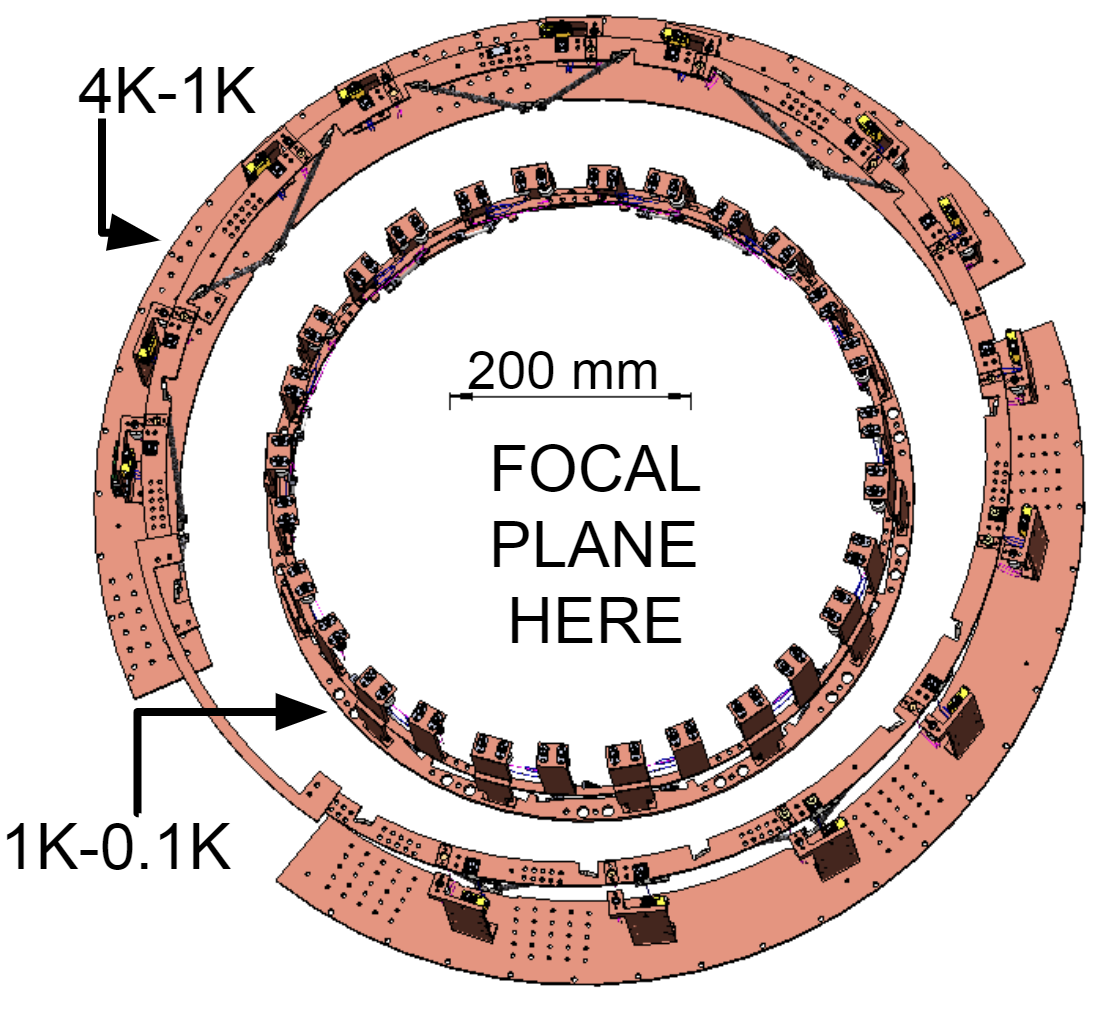} %SAT_annotated_cross_section.PNG
        \caption{SAT cold wiring design top view}
        \label{fig:SAT_design}
    \end{subfigure}
    \hfill
    \begin{subfigure}[b]{0.4\textwidth}
        \centering
        \includegraphics[width=\textwidth]{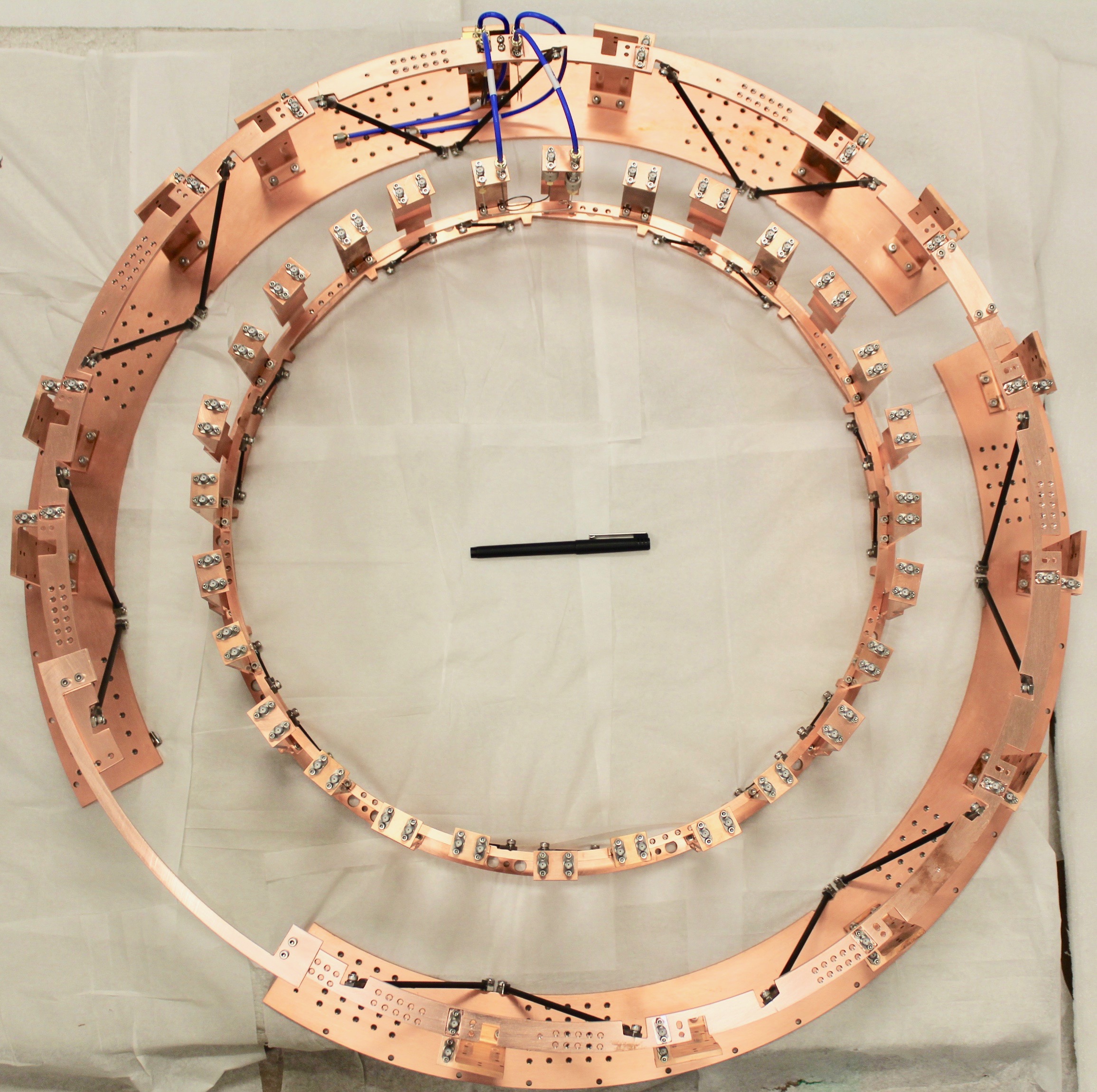} %SAT_1chain_pic.JPG
        \caption{Top view of SAT cold wiring during assembly}
        \label{fig:SAT_picture}
    \end{subfigure}
    \caption{(a) Single RF chain for SO readout. Attenuators are distributed to provide optimal drive power, while reducing equivalent noise temperature and reflections. Two stage cryogenic amplification is used for improved SNR and linearity. Semi-rigid coaxial cables of different materials, lengths and diameters are used for inter-thermal stage interconnects. DC-blocks provide thermal breaks between temperature stages. The Universal Readout Harness (URH) design is common to LATR and SATs. URH design in (b) and URH as pictured during assembly stage in (c). The design of cold readout wiring over 4 K to 0.1 K for the LATR (d) with wiring picture of a prototype in (e). The SAT cold wiring design top view in (f) with the hardware pictured in build stage show in (g). }\label{fig:cold_wiring}
\end{figure}

The wiring diagram for a single RF chain in SO is shown in Fig. \ref{fig:SO_wiring}, where a chain refers to the components and wiring that service one set of multiplexed detectors.  Attenuators are distributed over the various cryogenic stages of the receivers with the attenuation values chosen to effectively reduce noise temperature while providing optimal input power to drive the cold resonators. %and reducing reflections from impedance mismatch. 
Semi-rigid coaxial cables connect the inter-thermal stage RF components with cable material and lengths chosen to maximize signal-to-noise ratio (SNR) and reduce thermal loading. DC-blocks serve to provide thermal breaks. For improved SNR and linearity, cryogenic amplification is implemented in two stages in the output chain. For isothermal cabling we use hand-formable copper coaxial cables. %The wiring is design is modular. 
The Universal Readout Harness (URH) (see Figs. \ref{fig:URH_design} and \ref{fig:URH_picture}) is a module that is designed and is currently being populated and tested at Arizona State University (ASU), which includes all the wiring over 300 K to 4 K (excluding the 4K amplifiers that are mounted in the optics tubes of the LAT receiver and SATs). The URH design is common to both the LATR and the SATs. The wiring design from the URH at 4 K to the UMMs at 0.1 K in the LAT-Receiver (LATR) and the SATs are shown in Figs. \ref{fig:LATR_design}, \ref{fig:LATR_picture} and Figs. \ref{fig:SAT_design}, \ref{fig:SAT_picture} respectively.  These vary in cable lengths and routing due to the differences in the optics tubes of each, while conforming to the scheme presented in Fig. \ref{fig:SO_wiring}.

\section{Design Considerations}
\subsection{Linearity}\label{sec:linearity}
In the cryogenic readout circuit there are two primary nonlinear elements: the resonators and amplifiers. The superconducting microwave resonators have a nonlinear penetration depth, surface impedance \cite{Dahm1997_pendepthsurfznonlinearity}, and readout power heating \cite{deVisser2010_ReadPowerNonlinearity}. The Josephson junctions in the RF-SQUIDs also add non-linearity to the resonator circuits. All of these can result in resonator bifurcation and hysteresis \cite{Siddiqi2005_JJNonlinearity}. The output amplifier chain is another source of non-linearity with input power producing third order intermodulation products within the readout band. We plan to read out \bigO(1,000) bolometers on one transmission line. This means that the output amplifier chain sees \bigO(30~dB) higher power at its input than any single resonator. Due to this, the linearity requirements placed on the amplifiers are significantly higher than those placed on the resonators. 

To determine the linearity requirements on the amplifiers we first need to determine the optimal drive power for the resonators. Given a perfectly linear output amplifier chain the resonator drive power is optimized by minimizing readout noise. As the drive power is increased the TLS noise contribution decreases \cite{Gao2008_TLSvsP} until TLS noise saturates and there is no additional benefit in further increasing drive power. However driving resonators at high powers can result in increased noise by mechanisms such as resonator bifurcation \cite{Gaothesis} and other non linear-effects in the system. From measurements of similar devices from NIST \cite{mumux_cmb}, the expected optimum drive power for SO resonators is $\sim$ -70~dBm. 

As presented previously by Henderson \textit{et al.} \cite{smurf}, we will read SO $\mu$-mux resonators out with  SMURF electronics, which implements a frequency tracking algorithm in the on-board FPGA. Frequency tracking is crucial because it keeps the power transmitted to the amplifiers constant at a level lower than the drive power by the resonance depth. SO resonators are designed with an assumed $Q_i$ of 200,000 and targeting a fixed bandwidth (BW) of 100~kHz to keep Lorentzian crosstalk constant across the readout bandwidth. To achieve this the coupling quality factor ($Q_c$) is designed to vary between 50-133 $\times10^3$ over 4-8~GHz, such that the resonance depths $\sim Q_c/(Q_i+Q_c)$ vary between $\sim$ 8-14~dB. With an average designed resonance depth of $\sim$ 10~dB, the average power seen at the input of the amplifier for each tone is $\sim-$80~dBm and with \bigO(1,000) tones the total power transmitted to the first amplifier is $\sim-$50~dBm

This number sets the input third order intercept (IIP3) of the first amplifier. All successive amplifiers are chosen to have IIP3s lower than the output IP3 (OIP3) of the previous amplifier so that the first amplifier limits the linearity of the entire chain. Ideally, one would have an amplifier chain with linearity much larger than the expected signal, but practically gain, linearity, and noise temperature (as well as power dissipation and impedance matching) can not all be optimized simultaneously. At least one parameter must be sacrificed in order to improve the others. To maintain a low noise temperature and achieve a high IIP3 we choose a lower-gain first-stage amplifier at 4~K and add a second amplifier at 40~K to provide sufficient system gain. This 40~K amplifier  achieves higher linearity compared to the 4~K amplifier due to the lower gain and noise temperature. The final amplification chain summarized in table \ref{tab:ampchain} is expected to meet linearity requirements for the SO goal noise level of 45~\parthz, with 2000x multiplexing. Preliminary noise models suggest resonance depths greater than 6 dB with tone tracking to achieve this noise level. For the SO baseline noise level of 65~\parthz and with 1000x multiplexing, the same noise model requires the resonance depths to be greater than 3dB (also with tone tracking). %Any additional amplifiers after the second stage can be achieved by off-the-shelf, inexpensive, higher noise temperature, high-linearity amplifiers.

\begin{table}
\centering
\renewcommand{\arraystretch}{4.0}
    \begin{tabu} to \textwidth {|X[5,c,m]|X[4,c,m]|X[2,c,m]|X[1.2,c,m]|X[1.2,c,m]|X[1,c,m]|X[1.2,c,m]|}
 \hline
 \textbf{Expected Power in 1000 tones at Input to Amplifier}&\textbf{Amplifier Part Number}&\textbf{Physical Temperature}&\textbf{IIP3}&\textbf{Input T$_{\mathrm{Noise}}$}&\textbf{Gain}&\textbf{OIP3}\\ 
 \textbf{[dBm]}&&\textbf{[K]}&\textbf{[dBm]}&\textbf{[K]}&\textbf{[dB]}&\textbf{[dBm]}\\ 
 \hline
 -50&Low Noise Factory LNC4\_8LG&4&-30&2&24&-6\\ 
 \hline 
 -26 & Arizona State University SO 40K & 40 & 7 & 35 & 14 & 21\\
 \hline
  -14 & Minicircuits ZX60-83LN12+ & 300 & 13 & 100 & 20 & 33\\
 \hline
\end{tabu}
\caption{Summary of linearity and noise temperature of SO readout amplifiers.}
\label{tab:ampchain}
\end{table}
\vspace{-0.5mm}
\subsection{Reflections}\label{sec:reflections}
Reflections resulting from poor impedance matching between successive components in the RF chain cause frequency-dependent power coupling to the microwave transmission line of the $\mu$-mux wafer and consequently the superconducting resonators. Poor impedance matching and the resulting reflections can produce spectral features in addition to the frequency-dependent insertion loss of cables \cite{Pozar}. As described in Sec. (\ref{sec:linearity}) the drive power of resonators is optimized to reduce two-level system noise and non-linearities from resonator bifurcation. Although ideally we desire perfect impedance matching, in practice we set a threshold input return loss of all components to be better than $-15$~dB, by careful selection of cables and components that are well matched to a characteristic impedance of $Z_0=50~\Omega$. %Additionally, we introduce attenuators to minimize reflections and empirically verify that the drive power at the input of the $\mu$-mux wafer is at the desired level. 
Furthermore, to minimize degradation of SNR in the output RF chain due to poor impedance matching, we introduce a non-reciprocal device, such as a wide-band RF isolator, between the microwave launches on the  $\mu$-mux wafer and the first stage amplifier at $4~\rm{K}$. With suitable choice of components that have reasonably good impedance matching ($\sim-15$~dB) measurements from similar RF chains in test cryostats \cite{smurf} satisfy noise requirements discussed in Sec.~\ref{sec:noise}.
%After first stage amplification, we do not find the need for an additional isolator between amplifiers at $4~\rm{K}$ and $40~\rm{K}$ because the input reflections of the 40 K amplifier are much smaller than the 4 K amplifier.

\subsection{Thermal loading}\label{sec:thermal}
%TES bolometer noise and consequently instrument sensitivity is primarily driven by three components namely photon noise, thermal carrier noise, and readout noise\cite{Bolocalc}. Photon noise is determined by the telescope optics and we discuss readout noise in Sec.{\ref{sec:noise}}. Thermal carrier noise depends critically on the operating temperature of the bolometer and the thermal bath on which the superconducting film is suspended, and it is essential to maintain the temperature of the focal plane at stable sub-Kelvin temperatures. 
With the large cryostats optimized for high throughput employed for SO, a large fraction of the cooling budget is taken up by radiative optical loads, and conductive loading from mechanical supports \cite{SO_instrument}. We use semi-rigid coaxial\footnote{from COAX CO., LTD. \tt{http://www.coax.co.jp/en/}} cables with long lengths and small cross-sectional area to connect components at different temperatures to reduce conductive thermal loading, while providing robust electrical performance. For making inter-stage connections in the input RF chain we use Cupronickel (CuNi) semi-rigid coaxial cables (2.19 mm diameter from $300~\rm{K}$ down to $4~\rm{K}$ and 0.86 mm diameter from $4~\rm{K}$ to $0.1~\rm{K}$). To minimize losses in the output chain and hence avoid degradation of SNR we use superconducting Niobium Titanium (NbTi) cables of diameter 1.19 mm from the $\mu$-mux wafer at $0.1~\rm{K}$ to first stage amplification at $4~\rm{K}$. We also introduce inner-outer DC blocks between inter-stage wiring to provide thermal breaks. For interconnects between isothermal components we use 3.58 mm diameter hand-formable copper coaxial cables for ease of wiring over long lengths and tight spaces. Particular attention is paid to the heat-sinking of active components such as amplifiers as well as passive high-loss components such as attenuators. Mechanical structures on which readout components are mounted are designed to reduce heating from vibrational modes that result from telescope slew. We thus ensure that total thermal loading contribution from wiring, dissipation from readout components and associated mechanical structures are designed to be within the allocated readout cooling power specifications for the SATs and the LATR. For illustration the estimated readout thermal loading in one SAT is in table \ref{tab:thermal}. The simulated loading presented here fits well within allocated SO readout thermal loading budget and cryogenic validation of readout wiring is underway in test dewars. 
\begin{table}[htbp]
\centering
\renewcommand{\arraystretch}{2.0}
\begin{tabu} to  
\textwidth {| X[c,m] | X[c,m] | X[c,m] |}
 \hline
 \textbf{Nominal Stage Temperature} & \textbf{Total Cooling Power} & \textbf{Estimated Readout Thermal Loading}\\ 
 {\textbf{[K]}} & \textbf{[W]} & \textbf{[W]} \\
 \hline \hline
 40 & 55 & 2.4 \\ 
 \hline
 4 & 2 & 0.2\\
 \hline
 1 & $3\times10^{-2}$ & $4\times 10^{-5}$ \\
 \hline
 0.1 & $4\times10^{-4}$ & $11.6\times 10^{-6}$ \\
 \hline
\end{tabu}
\caption{Cooling power requirements for readout cabling and components between 300 K to 0.1 K. Thermal model includes conductive loading, microwave power dissipation from attenuation (including cables), and dissipation from amplifiers. The estimated loading from readout is a small fraction of the available cooling power at all thermal stages.}
\label{tab:thermal}
\end{table}

%\vspace{-1.5mm}
\subsection{Size}\label{sec:size}
A targeted multiplexing factor of \bigO(1,000) to readout \bigO(10,000) detectors results in running \bigO(10) cables through each optics tube of the LATR and each SAT. This necessitates maximizing packing density of the readout components and cables in the tight cryogenic volume of the receiver, designing wiring with low mechanical footprint. Semi-rigid coaxial cables are bent to include loops for strain relief, but with tight bend radii conforming to vendor specified minimum bend radii. The tight space constraints are best exemplified by the wiring designs shown in Figs. \ref{fig:URH_design}, \ref{fig:LATR_design}, and \ref{fig:SAT_design}. The SAT requires that the detector modules be removed without having to remove the majority of the readout wiring, which we achieved with the tight radial design shown in Fig. \ref{fig:SAT_design}. The LATR optics tubes is designed so that it can be removed without having to remove the majority of the readout wiring. It has a very short linear space behind the focal planes in the optics tubes ($\sim$5 cm from the top of the detector module to the 1 Kelvin magnetic shield), which we achieve with a very compact component mounting and coax routing design in the optics tubes shown in Fig. \ref{fig:LATR_design}.

\subsection{Noise}\label{sec:noise}
For our goal instrument configuration we require that the noise contributed by the readout does not increase the total detector white noise level by more than 5\%. In units of current in the SQUID input coil this corresponds to 45~\parthz. The main noise sources related to the RF chain are from referred thermal noise due to loss in the input chain and the intrinsic noise temperature of the output amplifiers. The Friis formula \cite{Pozar} is used to refer both of these noise sources to effective noise power (or noise temperature) at the resonator. We then relate the noise power at the feed-line to an effective flux in the SQUID and then to a current through the TES by the mutual inductance between the SQUID and the TES. The steps of this referral from noise power at the feedline to noise in the SQUID is discussed in theses by Mates \cite{Matesthesis} and Gao \cite{Gaothesis}. These noise sources are summarized in table (\ref{tab:noise}); note that there are additional noise sources associated with the flux ramp input lines, TES input lines, and resonators themselves which we did not consider as part of the RF coaxial component chain noise sources. The anticipated contributions from these key noise sources associated with the RF chain design are small compared to the total noise budget. Measurements with $\sim$500 multiplexed resonators  have demonstrated that the readout noise (DAC and amplifier chain noise plus resonator noise including TLS) is below the goal noise level \cite{smurf}. A future publication will present a detailed readout noise analysis including analytic calculations and measurements from lab tests.
\begin{table}[htbp]
\centering
\renewcommand{\arraystretch}{1.5}    \begin{tabu} to \textwidth {| X[c,m] | X[c,m] | X[c,m] |}
 \hline
 \textbf{Component} & \textbf{Noise} & \textbf{Effective T$_{\mathrm{Noise}}$} \\ 
 &\textbf{[}$\mathrm{\textbf{pA}}\textbf{/}\sqrt{\mathrm{\textbf{Hz}}}$\textbf{]}&\textbf{[K]}\\
 \hline \hline
 SMuRF Tone Generation & 15.6 & 20\\ 
 \hline
 Amplifier Chain & 6.2 & 3.4\\
 \hline
 Attenuator and Coax & 0.9 & 0.2\\
 Thermal Noise & & \\
 \hline
\end{tabu}
\caption{Summary of noise sources associated with the RF chain.}%The total allowable readout noise contribution is 45 \parthz.
\label{tab:noise}
\end{table}

\section{Conclusions}
The RF cryogenic chain from the 300K vacuum interface of SO receivers to the superconducting resonator chips has been designed based on thermal, linearity, noise and mechanical considerations. This design is currently being evaluated in test cryostats, and will be implemented in all Simons Observatory telescope receivers.
\vspace{-4.5mm}
\begin{acknowledgements}
This work was supported in part by a grant from the Simons Foundation (Award \# 457687.)
\end{acknowledgements}
\vspace{-2.5mm}
\bibliographystyle{spmpsci} %abbrv
\bibliography{LTD18.bib} 
\end{document}